\documentclass[notitlepage,prd,twocolumn,showpacs,preprintnumbers,nofootinbib,superscriptaddress,floatfix,tightenlines, longbibliography]{revtex4-1}

\usepackage{graphicx}
\usepackage{xcolor}

\usepackage{amsmath}
\usepackage{amssymb}
\usepackage{hyperref}
\usepackage{ulem}

\newcommand{\vb}{\vec}
\renewcommand{\vec}[1]{\mathrm{\mathbf{#1}}}
\newcommand{\dd}[2][]{\mathrm d^{#1}{#2}\,}

\newcommand{\nX}{n_X}

\newcommand{\Conetwo}{\mathcal {C}^{1\leftrightarrow 2}}
\newcommand{\Ctwotwo}{\mathcal{ C}^{2\leftrightarrow 2}}

\newcommand{\Nc}{N_C}
\newcommand{\vbphat}{\hat{\vb p}}
\newcommand{\qperp}{q_\perp}
\newcommand{\pmin}{p_{\mathrm{min}}}
\newcommand{\pmax}{p_{\mathrm{max}}}

\begin{document}
\title{Solving the QCD effective kinetic theory with neural networks}

\author{S. Barrera Cabodevila}
\affiliation{Instituto Galego de Física de Altas Enerxías IGFAE,
Universidade de Santiago de Compostela, E-15782 Galicia-Spain}

\author{A.~Kurkela} 
\affiliation{Faculty of Science and Technology, University of Stavanger, 4036 Stavanger, Norway}

\author{F.~Lindenbauer} 
\affiliation{Institute for Theoretical Physics, TU Wien, Wiedner Hauptstrasse 8-10, 1040 Vienna,
Austria}

\begin{abstract}
Event-by-event QCD kinetic theory simulations are hindered by the large numerical cost of evaluating the high-dimensional collision integral in the Boltzmann equation. In this work, we show that a neural network can be used to obtain an accurate estimate of the collision integral in a fraction of the time required for the ordinary Monte Carlo evaluation of the integral. We demonstrate that for isotropic and anisotropic distribution functions, the network accurately predicts the time evolution of the distribution function, which we verify by performing traditional evaluations of the collision integral and comparing several moments of the distribution function. This work sets the stage for an event-by-event modeling of the pre-equilibrium initial stages in heavy-ion collisions.

\end{abstract}

\maketitle

\section{Introduction}A short-lived fireball of strongly interacting matter is created in ultra-relativistic nuclear collisions performed at RHIC and the LHC \cite{Harris:2024aov}. While it may be that these systems thermalize locally in central collisions of large nuclei, admitting approximating the time evolution using a fluid-dynamic description, the evolution of peripheral collisions and collisions of light ions likely remains far from equilibrium throughout the entire evolution \cite{ExTrEMe:2023nhy, Romatschke:2016hle}. Even in the case of central heavy-ion collisions, the system is bound to be far from equilibrium at early times, at late times, and all times near the edge of the fireball. 
Nevertheless, the fluid-dynamic description has enjoyed great phenomenological success in describing the heavy-ion collisions \cite{Nijs:2020roc, Bernhard:2016tnd, Dusling:2007gi}. However, increased accuracy requirements in heavy-ion collisions and, in particular, the qualitative questions arising from observed signatures of collectivity in proton-nucleus and proton-proton collisions call for a full microscopic far-from-equilibrium description of the whole collision or a part of its time evolution \cite{Kurkela:2018wud, Gale:2021emg,Liu:2015nwa}. These systems pose a challenge for the phenomenological modeling that---if solved---may give an empirical inroad to test theoretical models of thermalization and hydrodynamization. 

The QCD effective kinetic theory (EKT) is one possible framework to solve for the far-from-equilibrium evolution \cite{Arnold:2002zm, Kurkela:2015qoa}. It becomes leading-order accurate in the theoretically clean limit of asymptotically large center-of-mass energies, and it can be extrapolated to the physical coupling values to provide a microscopic model of the far-from-equilibrium evolution that is firmly based in QCD. Compared to hydrodynamics, which follows only the conserved currents, the kinetic theory describes the entire distribution function of the partons, making numerical solutions to the effective theory significantly more challenging. This has limited the application of the EKT to simple systems exhibiting a large degree of symmetry \cite{AbraaoYork:2014hbk, Kurkela:2014tea, Kurkela:2015qoa, Kurkela:2018vqr, Kurkela:2018xxd, Kurkela:2018oqw, Kurkela:2021ctp}. To date, a full event-by-event simulation with realistic ensembles of fluctuating initial conditions has remained prohibitively expensive. On the other hand, significant progress in event-by-event simulations has taken place with kinetic theories with simpler collision kernels \cite{Kurkela:2020wwb,Taghavi:2025xhl,Ambrus:2024hks}.

The main challenge of the EKT simulation is the evaluation of the collision integral, $v^\mu \partial_\mu  f(x^\mu, \vec p) = C[f](\vec p)$, which requires an evaluation of a distribution-dependent eight-dimensional integral at all the space-time points
spanned by the simulation. While it is feasible---albeit time-consuming ---to directly evaluate the collision integral to a good accuracy at simulation time using the Monte Carlo methods developed in \cite{AbraaoYork:2014hbk, Kurkela:2015qoa} and used widely in the literature \cite{Kurkela:2018oqw, Du:2020zqg}, alternative and more efficient strategies would be desirable. Here we suggest such an alternative strategy separating the space-time evolution from the evaluation of the collision kernel. 

Crucially, while the collision kernel is non-local in the momentum space, it remains local in coordinate space. Then, at any given space-time point, the collision kernel is a deterministic, non-linear map from the distribution function to its time derivative, which can be, in principle, fitted or parametrized with a sufficiently versatile fitting function. Here, we develop such a fit by training a neural network with a large amount of training data obtained by using the traditional Monte Carlo evaluation of the collision kernel \cite{Xiao_2021,Xiao_2023}. The benefit of this approach is that once the neural-network model of the collision kernel is trained, it can be used very efficiently in any time evolution of complicated event-by-event time-evolutions instead of spending large amounts of computing resources on repeated evaluations of distribution functions that closely resemble each other, both across different space-time points and across events. In fact, it has been observed that the phenomenon of non-equilibrium attractors \cite{Heller:2015dha, Kurkela:2019set, Almaalol:2020rnu, Boguslavski:2023jvg} reduces the relevant space of relevant distribution functions, greatly simplifying the functional space needed to cover when training the neural network when applied to expanding kinetic theory simulations.

This work sets the stage for performing QCD kinetic theory simulations with inhomogeneous backgrounds, which has so far only been achieved using simple approximations for the collision integrals \cite{Kurkela:2019kip, Kurkela:2018qeb, Kurkela:2018ygx, Ambrus:2022koq, Ambrus:2022qya, Taghavi:2025xhl,Ambrus:2024hks}.

The structure of the paper is as follows. In Section \ref{sec:collision-kernel}, we discuss how the collision kernel can be obtained using the Monte Carlo method, and how the conformal symmetry of QCD kinetic theory restricts our training data set. In Section \ref{sec:isotropic-systems}, we provide results for isotropic systems, and in Section \ref{sec:anisotropic-distributions} for anisotropic distribution functions. We conclude in Section \ref{sec:conclusions}. Appendix \ref{sec:collision-kernels} discusses numerical details on the Monte Carlo method of obtaining the collision kernels, and Appendix \ref{app:restframe} describes in detail how we perform the transformation to the rest frame.

\begin{figure}
    \centering
    \includegraphics[width=\linewidth]{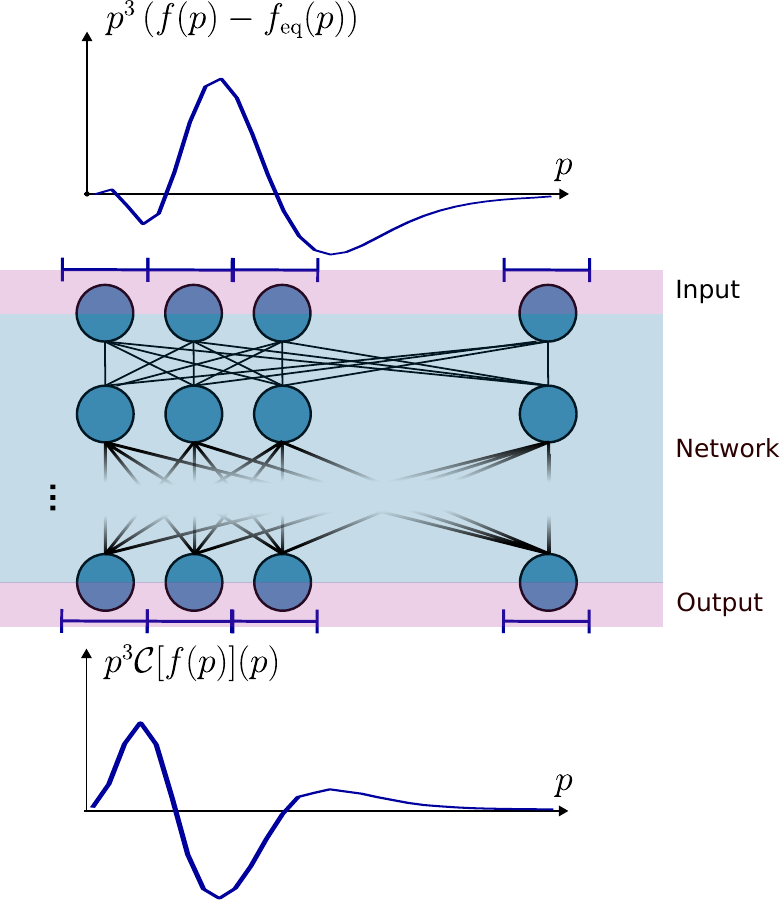}
    \caption{Schematic overview of the use case of our neural network. 
    It takes the distribution function (with thermal equilibrium subtracted) as input and provides a prediction for the collision kernel $\mathcal C$.
    }
    \label{fig:network}
\end{figure}

\section{The collision kernel \label{sec:collision-kernel}}
\subsection{Collision kernel in EKT}
The QCD effective kinetic theory (EKT) \cite{Arnold:2002zm} is based on the Boltzmann equation for the space-time and momentum-dependent distribution function for on-shell partons $f(x^{\mu}, \vec p )$,
\begin{align}
 v^\mu \partial_\mu f(x^{\mu}, {\vb p} ) =-\mathcal{C}[f(x^\mu, \vb p)], \label{eq:boltzmann_equation}.
\end{align}
Here, the velocity is $v^\mu = p^\mu/p^0$. The left-hand side accounts for the free-streaming space-time evolution and is local in $\vec p$-space but is not local in $\vb x$-space in the sense that it connects neighboring points by derivatives. The collision kernel $\mathcal{C}[f(x^\mu, \vec p)]$ of the right-hand side accounts for the scattering between the partons, and is strictly local in $\vec x$-space but nonlocal in $\vb p$-space.  It consists of elastic $2 \leftrightarrow 2$--scattering and medium-induced (effective) collinear $1\leftrightarrow 2$ splitting/merging processes
\begin{equation}
\mathcal{C}[f] = \Ctwotwo[f] + \Conetwo[f], \label{eq:collisionkernel-splitting}
\end{equation}
both of which are local in space-time $x^\mu$ but non-local in momentum space $\vec p$. In this exploratory work, we will consider only gluons for simplicity. For the inclusion of quarks in EKT, see \cite{Kurkela:2018oqw,Du:2020dvp}.

The elastic collision term consists of a gain and loss term and is given by
\begin{align}
\Ctwotwo[f(\vb p)]&=\frac{1}{4|\vb{p}|\nu}\int_{\vb{kp'k'}}\left|\mathcal M(\vb{p},\vb{k};\vb{p'},\vb{k'})\right|^2 \nonumber\\
&\quad\times(2\pi)^4\delta^4(P+K-P'-K')\label{eq:c22_first}\\
&~~~\times\Big\{f(\vb p)f(\vb k)\left[1 + f(\vb p')\right]\left[1+ f(\vb k')\right]\nonumber\\
&~~\quad - f(\vb p')f(\vb k')\left[1+ f(\vb p)\right]\left[1 + f(\vb k)\right]\Big\} \nonumber ,
\end{align}
where $\int_{\vb k}=\int\frac{\dd[3]{\vb k}}{(2\pi)^3 2|\vb k|}$ denotes the Lorentz-invarant integration measure for on-shell momenta ($K^0=|\vb k|$), and $\nu=2d_A{=2(N_c^2-1)}$ counts the number of degrees of freedom for gluons. For simplicity, we have suppressed the space-time coordinate $x^\mu$ in all distribution functions, since it is local.
For the matrix element we use the leading-order pQCD $gg-gg$ scattering matrix element summed over all incoming and outgoing spin and color degrees of freedom. We include medium effects by screening the divergent $t$ and $u$ channel propagators using the isotropic HTL (isoHTL) propagators for the internal soft gluon lines \cite{Arnold:2002zm, Boguslavski:2024kbd}. We use the 't Hooft coupling $\lambda=g^2\Nc=10$ throughout this paper.

The inelastic collision term accounts for splitting and merging processes, and has a similar structure as \eqref{eq:c22_first} except only containing three in- and outgoing particles.
In its symmetrized form, it is given by
\begin{align}
    \begin{split}\label{eq:c12}
        \Conetwo[f(\vb {\tilde p})]&=
        \frac{(2\pi)^3}{4\pi\tilde p^2}\frac{1}{\nu}\int_0^\infty \dd{p}\int_0^{p/2}\dd{k'}4\pi\gamma^{p}_{p',k'}\\
        &\quad\times\Big\{f(\vb p)(1+f(p'\vbphat))(1+f(k'\vbphat))\\
        &\qquad\quad-f(p'\vbphat)f(k'\vbphat)(1+f(\vb p))\Big\}\\
        &\times\left[\delta(\tilde p-p)-\delta(\tilde p-p')-\delta(\tilde p-k')\right]
    \end{split}
\end{align}
With only two momenta and strictly collinear splitting in the direction of $\vb p$ (formalized by the unit vector $\vbphat=\vb p/p$), its kinematics is simpler than for the elastic collision term. However, the effective matrix elements for the splitting/merging processes $\gamma$ are more complicated since they need to account for the quantum-mechanical formation time of the splitting-merging process, and effectively interpolate between the Bethe-Heitler and LPM rates \cite{Arnold:2002ja}. We describe the collision terms in more detail in the Appendix \ref{sec:collision-kernels}.

\subsection{Collision Kernel as a machine-learning problem}

More generally, we can view the collision kernel as a map of the space of distribution functions to their time derivatives {$\mathcal C:f\mapsto (\partial_t f)_{\mathrm{coll}}$}. After discretizing momentum space $\vb p= (p_i, \cos\theta_j, \phi_k)$, we can represent the distribution function at every space-time point as a $\nX$-dimensional vector, $f(\vb p)\in \mathbb R^{\nX}$. The collision kernel is then a nonlinear map
 \begin{equation}
 \mathcal{C}: \mathbb{R}^{\nX} \to \mathbb{R}^{\nX},
 \end{equation}
 with $n_X$ being the number of points in the discrete momentum space. Modeling such a map is a task well-suited to an artificial neural network \cite{HORNIK1989359, Mehta:2018dln}.

The collision kernel can be computed using Monte Carlo integration techniques, as detailed in Appendix \ref{sec:collision-kernels}. However, this is costly, and many of the distribution functions closely resemble each other. Therefore, in a fully 3+3+1D system (corresponding to a distribution function $f(\vb x,\vb p,t)$), we need to repeatedly obtain the collision kernel for very similar distribution functions. However, since the collision kernels are local in $\vec{x}$-space, their calculation only involves the value of the distribution function in the current cell of the spatial grid. In this work, we demonstrate that we can compute these collision kernels efficiently with machine learning techniques. For this reason, we don't include the spatial dependence in our results and we leave the full 3+3+1D simulation for the future.

Our strategy in the following is to train a neural network to perform this map. We generate training data using numerical evaluations of the collision kernel using Monte Carlo integration, which we describe in more detail in Appendix \ref{sec:collision-kernels}.
In order to reduce the needed amount of training data, we utilize symmetries in the collision kernel. One way to use symmetries in machine learning frameworks is to use network architectures that by construction preserve symmetries \cite{2017arXiv171005468K, 2016arXiv160207576C, 2020arXiv200405154E, Favoni:2020reg, rath2024boosting, Gerken:2021sla}. We proceed differently here. Instead of using a network architecture that manifestly preserves equivariance in the symmetry transformation, we perform the symmetry transformation on the input data and train the network only on a subset.

To illustrate, for instance, we have the freedom to choose a coordinate system in momentum space, and, in particular, can perform arbitrary rotations to this coordinate system. If we rotate the input distribution, the collision kernel will be rotated by the same amount. We can now choose to always rotate the input distribution such that the most anisotropic direction (which is to be made more precise below) is the $z$-direction, and then train the network only on the subset of data where this is the case. Afterward, we rotate back.

We will discuss the precise symmetry transformations used here in the next subsection.

\subsection{Conformal symmetry of the EKT}
\label{sec:conformalsymmetry}
The collision kernel map $\mathcal C$ naturally respects certain symmetries.
For any given distribution function $f(\vb p,\vb x, t)$, at every spacetime point $(\vb x,t)$, there exists a rest frame in which the energy-momentum tensor $T^{\mu\nu}$ is diagonal, 
\begin{align}
    &T^{\mu\nu}(\vb x,t)=\nu\int\frac{\dd[3]{\vec p}}{(2\pi)^3}\frac{p^\mu p^\nu}{p^0} f(\vec p,\vb x, t)\label{eq:tmunu}\\
    &=\varepsilon(\vb x,t) u^\mu_0(\vb x,t) u^\nu_0(\vb x,t) 
    + \sum_i P_i(\vb x,t) u^\mu_i(\vb x,t) u^\nu_i(\vb x,t),\nonumber
\end{align}
with $\{u_i^\mu \}$ the eigenvectors of $T^{\mu\nu}$, and its diagonals are given by the energy density $\varepsilon$ and the pressures in different directions $P_i$.
We train our network only on distribution functions in the rest frame, obtained by a suitable Lorentz transformation as discussed in Appendix \ref{app:restframe}.
Additionally, for an anisotropic system, we also rotate our coordinate system to enforce $P_z< P_y <P_x$.

Furthermore, since we treat QCD in the conformal weak-coupling regime, neglecting the effects from running coupling, the only dimensionful scale is set by the energy density $\varepsilon$.
From this, we can define the instantaneous and local target equilibrium temperature (Landau matching)
\begin{align}
    T(\vb x, t)=\left(\frac{30\,\varepsilon(\vb x,t)}{\nu\pi^2}\right)^{1/4},
\end{align}
where $\nu$ is the number of degrees of freedom.
Since the collision kernel $\mathcal C$ is local in spacetime, we will not carry the spacetime dependence $(\vb x,t)$ explicitly.

Therefore, it is enough to train our neural network for one specific temperature (since every distribution function corresponding to another temperature can be rescaled) and in the rest frame with a specific orientation, which greatly reduces the number of input data needed.

In our implementation of the collision kernel,
we work with dimensionless quantities where the temperature is scaled out.

\subsection{Discretization}
We discretize the distribution function $f(\vb p)$ using the \textit{discrete-momentum method} introduced and described in Ref.~\cite{AbraaoYork:2014hbk}. Instead of the distribution function $f(\vb p)$ itself, we store the number density per bin
\begin{align}
    n_{ijk}=\int\frac{\dd[3]{\vb p}}{(2\pi)^3}f(\vb p) w_i(p)  \tilde w_j(\cos\theta)\hat w_k(\phi).\label{eq:def_nijk}
\end{align}
Here, the piecewise linear wedge functions $w$ are defined via
\begin{align}
    w_i(p)=\begin{cases}
        \frac{p-p_{i-1}}{p_i-p_{i-1}}, & p_{i-1} < p < p_i\\
        \frac{p_{i+1}-p}{p_{i+1}-p_{i}}, & p_{i} < p < p_{i+1}\\
        0& \text{else,}
    \end{cases}\label{eq:wedge-functions}
\end{align}
and the boundary bins are constructed as $w_0(p)=(p_1-p)/(p_1-p_0)$ for $p_0 < p<p_1$ and zero otherwise, and $w_n(p)=(p-p_{n-1})/(p_n-p_{n-1})$ for $p_{n-1} <p<p_n$ and zero otherwise.
The grid for the polar ($\cos\theta$) and azimuthal ($\phi$) angles is constructed similarly, with the exception that the polar angle grid is periodic, and thus there are no boundary bins.
At any given momentum $\vb p$, the distribution function $f(\vb p)$ is encoded in the neighbouring grid points $(p_i, \cos\theta_j,\phi_k)$. At the grid points themselves, the relation is approximately
\begin{align}
		f(p_i,\cos\theta_j,\phi_k)\approx\frac{(2\pi)^3n_{ijk}}{ p_i^2 \Delta V^p_i\Delta V^\theta_j\Delta V^\phi_k},
	\end{align}
where the volume factors are defined via
 \begin{align}
     \Delta V_i^p=\int_{-\infty}^\infty\dd{x}w_i(x).
 \end{align}
 Between the grid points, $f(\vb p)$ is found by linearly interpolating the neighboring results. Important discretization parameters are the grid boundaries for the $p$-grid, $\pmin$, and $\pmax$. In the presented results, we always choose $\pmax=15T$. However, the $\pmin$ value is different for the isotropic and anisotropic cases. We set $\pmin=0.01T$ in the former and $\pmin=0.1T$ in the latter.

\subsection{Data preprocessing and the coarse grid}
To generate training data using the Monte Carlo method, a fine grid with typically a large number of points per dimension must be employed, as the method of \cite{AbraaoYork:2014hbk} suffers from numerical instabilities during time evolution--—instabilities that are mitigated by finer grid spacing. These instabilities can arise when the collision kernel is evaluated between grid points where one is overoccupied and the other underoccupied.
However, despite the fine grid needed for the Monte Carlo method, the physical information can often be reduced to a small number of parameters, e.g., 
peak positions, width, and weight; or, more physically, moments of the distribution functions, which are used, for instance, in hydrodynamical models \cite{Denicol:2012cn}. Hence, some of the information stored in the fine grid in the Monte Carlo method is redundant, and for a neural network it may be sufficient to use a dimensionally reduced input data set.
In particular, for the reasonable accuracy needed for phenomenological applications, a very fine grid is also not needed.

Therefore, in this subsection, we introduce a coarse grid on which the input data for the neural network is stored.
In addition to the dimensional reduction, this coarse grid has another convenient feature: It reduces the noise coming from the Monte Carlo integration used in the Monte Carlo method for obtaining the collision kernel.

We will now describe how to efficiently transform from the fine grid to a coarser grid.
The system is initially described in terms of the number moments $n_{ijk}$ defined by \eqref{eq:def_nijk} on the fine grid. We can obtain the distribution function on a different grid as follows.
Representing the distribution function in terms of the old basis,
\begin{align}
    f(\vb p)\approx \sum_{ijk}\frac{(2\pi)^3n_{ijk}}{p_i^2 \Delta V_i^p\Delta V_j^\theta\Delta V_k^\phi}w_i(p)\tilde w_j(\cos\theta)\hat w_k(\phi),
\end{align}
we may obtain the representation in the coarse grid $n_{\alpha\beta\gamma}$ via the matrix multiplication
\begin{align}
    n_{\alpha\beta\gamma}=\sum_{ijk}\Omega_{\alpha\beta\gamma ijk}n_{ijk}
\end{align}
with
\begin{align}
\begin{split}
   \Omega_{\alpha\beta\gamma ijk}&=\int_0^\infty\dd{p} \frac{w_i(p)\omega_\alpha(p)}{\Delta V_i^p}
    \int_{-1}^1\!\!\dd{\cos\theta}\frac{\tilde w_j(\cos\theta)\tilde\omega_\beta(\cos\theta)}{\Delta V^\theta}\\ &\times\int_0^{2\pi}\!\!\dd\phi \frac{\hat w_k(\phi)\hat\omega_\gamma(\phi)}{\Delta V_k^\phi},
    \end{split}
\end{align}
where the $\omega$ functions are the corresponding wedge functions in the new grid. Note that the integrals are not nested, so we can decompose the full matrix as the tensor product of the matrix for each of the dimensions $\Omega_{\alpha\beta\gamma ijk} = \Omega_{\alpha i} \Omega_{\beta j}  \Omega_{\gamma k} $.

It should be noted that this matrix is not quadratic for different grid sizes and, as such, is not invertible (which reflects the loss of information in going from a fine to a coarse grid). 
Importantly, this matrix is independent of the form of $f(\vb p)$, and thus only needs to be computed once for every specific grid change.
This coarse grid $n_{\alpha\beta\gamma}$ will be used for the training of the neural network: Both the distribution function and collision kernel will lie in the coarse grid, even though the original data has been computed in a finer grid.

\section{Neural network model of the EKT: isotropic systems\label{sec:isotropic-systems}}
Having discussed how we use symmetries to reduce the amount of training data, we now turn to discuss how we generate training data in the isotropic case.
\subsection{Generation of training data, choice of initial conditions}
\begin{figure*}
    \centerline{
        \includegraphics[width=0.5\linewidth]{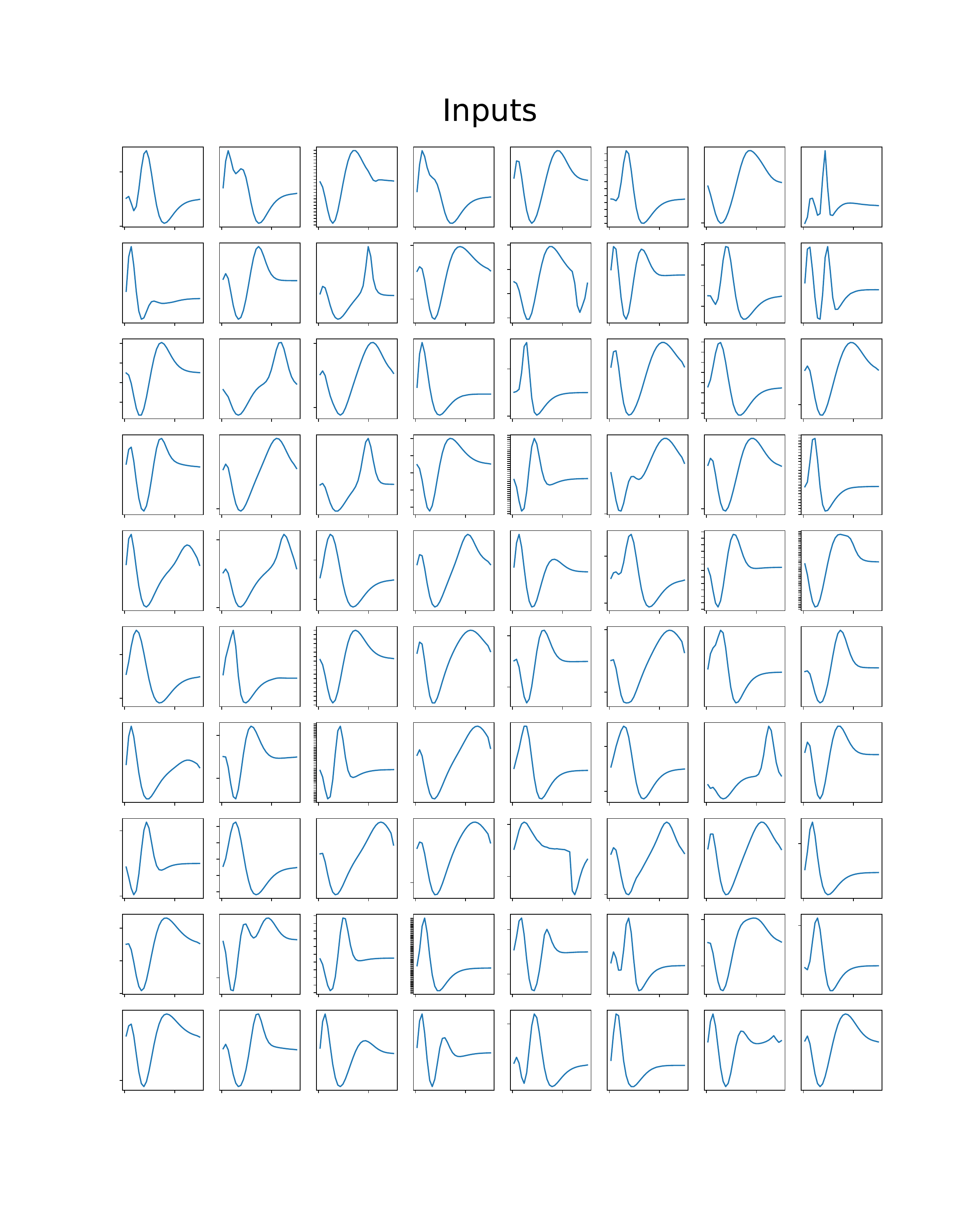}
        \includegraphics[width=0.5\linewidth]{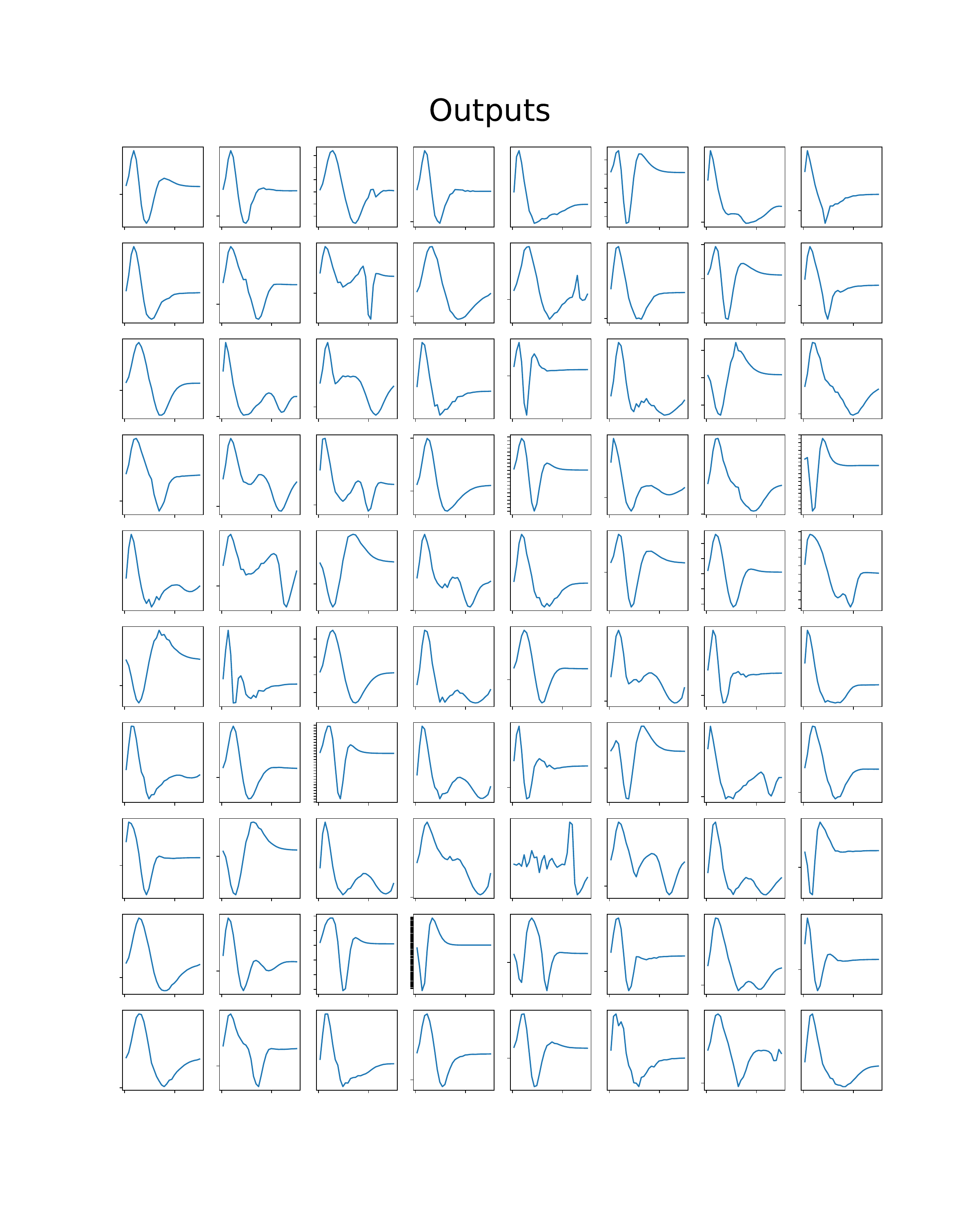}
    }
    \caption{Randomly chosen samples of our training data for the isotropic case. On the left, we show the input data $p^3(f(p) - f_{\mathrm{eq}}(p))$
    and on the right, we show the sum of the collision kernels $p^3C(f[p])$. The tick marks in the $y$-axis have a spacing $8\times 10^{-4}T^3$ in the inputs and $2.8\times 10^{-3}~T^4$ for the outputs. Panels with a larger number of tick marks correspond to distributions that are further from equilibrium.
    }
    \label{fig:dataset}
\end{figure*}
For training the neural network, we need to obtain a variety of training data that
accurately represents the input parameter space we wish to cover. Several considerations must be taken into account in order to construct an
exhaustive
dataset.

To create the training dataset, we consider several classes of distribution functions. For each class, we generate a sufficient amount of different distribution functions, obtain the collision kernel via the Monte Carlo method, and use this input-output data as part of our training set.

The first class consists of perturbations over thermal equilibrium.
This is motivated by the physical condition that a thermal system should remain thermal, and the network should learn that small perturbations around this equilibrium state generate collision kernels that lead back to equilibrium.

For this class, we start with a thermal distribution function,
\begin{align}
    f(p{/T})=\frac{1}{e^{p{/T}}-1},
\end{align}
and add a Gaussian peak
\begin{align}
    a_0e^{-s_0( p-p_0)^2{/T^2}}.
\end{align}
Naturally, this changes the energy density. We then remove energy density by ``adding a dip'', i.e., multiplying the resulting distribution function 
by a function $g(p)$ with
\begin{align}
    g(p{/T})=\begin{cases}
        1 & \text{almost everywhere}\\
        <1 & \text{for a very small region}
    \end{cases}
\end{align}
such that the energy density of the resulting system \eqref{eq:tmunu} corresponds to a thermal system with unit temperature.
We choose $g(p)$ of the form
\begin{align}
    g(p{/T})=1-a_2e^{-s_2(p-p_2)^2{/T^2}},
\end{align}
with $a_2$, $p_2$, and $s_2$ chosen such that the total energy is conserved, with $0<a_2 < 1$.

We complement this static class of data with a dataset obtained from different time evolutions. As initial conditions, we choose either
{\it far from equilibrium initial condition}, where we place several Gaussian peaks with random mean values, variance, and amplitude on the grid. Then, we normalize the sum of all of them. Another set of initial conditions is given by a 
{\it thermal background with perturbations}. There, we additionally add a thermal distribution.
    We ensure that we have a sufficient amount of training data, which is also close to equilibrium.
With these initial conditions, we obtain the collision kernel using the Monte Carlo method and perform the time evolution using simple Euler steps, where we regulate the step size using Heun's method.
For each simulation, we save several sets of distribution functions and corresponding collision kernels to be used as our training data.

A representative sample of the near- and far-from-equilibrium distribution functions and their corresponding collision kernels in the training data set is displayed in Fig.~\ref{fig:dataset}. Note that, instead of using the distribution function and its corresponding collision kernel as input and output, we choose $p^3(f - f_{\mathrm{eq}})$ as the former and $p^3C$ as the latter. We perform the multiplication by $p^3$ because we have observed improved energy conservation. Similarly, for the input data, we subtract the equilibrium distribution to facilitate the convergence towards the thermal fixed point.

\subsection{Hyperparameter optimization and uncertainty quantification\label{sec:networkarchitecture}}
Many different neural network architectures have been described and proposed in the literature, which are designed for different tasks and purposes.
Here, we use a feed-forward neural network with linear layers and ReLU functions as activation functions to obtain the collision kernel\footnote{ More precisely, if the output is modeled with a network of $N$ linear layers, it is obtained as $\vec{y} = L_N(L_{N-1}(\cdots(L_1(\vec{x}))))$. Here, $\vec{x}$ is the input array and $L_i$ the $i$-th layer of the network. Therefore, the output of the $M$-th layer is computed by taking the output of the previous layer and applying the activation function: $y^M_i = \sigma \left(\sum_jW_{ij}y^{M-1}_j + b_i \right)$. In our case, we use a ReLU function as activation, which is defined as $\sigma(x) = \max (0, x)$.}.
For the precise architecture, i.e., number of layers and nodes per layer, we use an automated optimization framework,
Ray Tune \cite{liaw2018tune}, to find the optimal set of hyperparameters.
This hyperparameter tuning has the freedom to explore models with up to 5 layers; each of these can have 16, 32, 64, 128, 192, 256, 384, or 512 neurons.

We allow Ray Tune to choose not only the architecture of our neural network but also the learning rate\footnote{The learning rate is the parameter used to quantify how much the gradient descent algorithm modifies the free parameters of the artificial neural network in each step of its training. Choosing an appropriate value is crucial to obtain a successful fit.} as an additional hyperparameter. The architectures and learning rates with the best performance for each collision kernel are presented in Table~\ref{tab:hyperparameters}.
This tuning is performed for both collision kernels separately.

\begin{table}
    \centering
    \begin{ruledtabular}
    \begin{tabular}{c c c}
        Kernel & Architecture (nodes per layer) & Learning rate \\
        \hline
        $\Conetwo$ & 512 & 0.001312 \\
        $\Ctwotwo$ & 384/128/192 & 0.000499 \\
    \end{tabular}
    \end{ruledtabular}
    \caption{Best hyperparameters found for each collision kernels.}
    \label{tab:hyperparameters}
\end{table}

\begin{figure}
    \centering
    \includegraphics[width=0.9\linewidth]{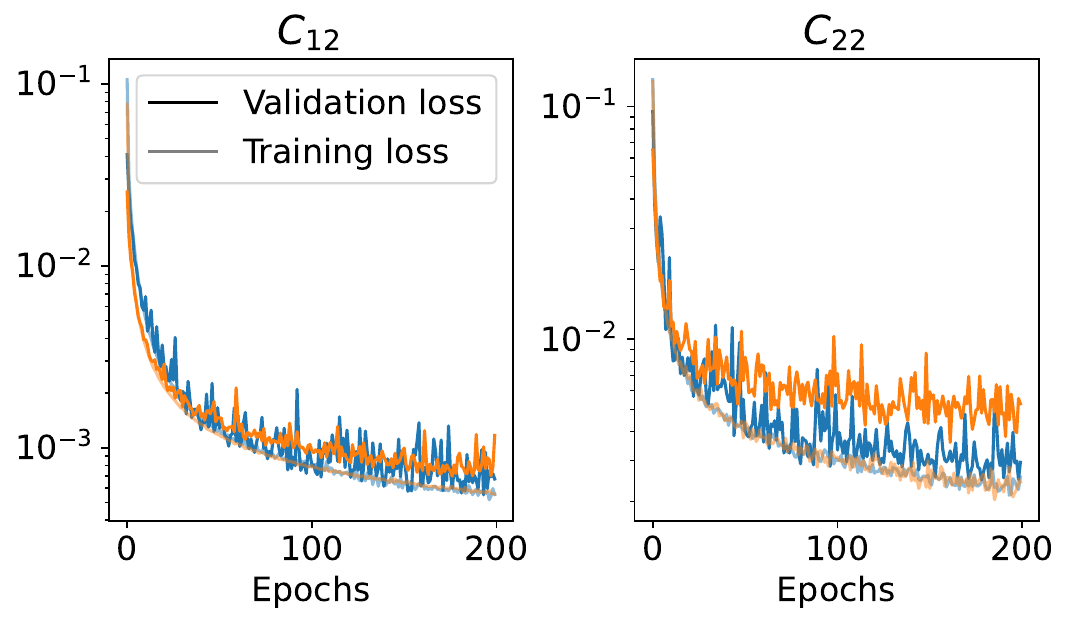}
    \caption{Training (lighter) and validation (darker) loss during each epoch of the training. The left panel shows the best (blue) and worst (orange) trainings for the $C_{12}$ kernel. The right panel shows the same but for $C_{22}$.}
    \label{fig:loss_1d}
\end{figure}

Once the Ray Tune has set the optimal hyperparameters (which involves training for a limited number of epochs) for each of the neural networks, the ten best networks are trained more extensively. 
For the training, we split the available training data randomly into training (80\%) and validation (20\%) subsets.
During the training, we minimize the loss function $\mathfrak L$, which is given by the mean squared error between the predicted and true (Monte Carlo) output values,\footnote{The discretized collision kernel $\mathcal C_{ijk}$ is defined analogously to the distribution function \eqref{eq:def_nijk}.}
\begin{align}
    \mathfrak L=\frac{1}{n_X}\sum_{ijk}\left[(p\mathcal C^{\mathrm{truth}})_{ijk}-(p\mathcal C^{\mathrm{NN}})_{ijk}\right]^2.
\end{align}
In Figure \ref{fig:loss_1d}, we show the best and worst training and validation losses as a function of the epoch of the training for each of the two collision kernels. It decreases rapidly at initial times and quickly approaches a plateau region. The best fit is given by the step at which the validation loss is minimal. 

After the hyperparameter tuning is done, we keep the best-performing 10 networks. This is to provide an error estimate for the network predictions. Since these networks differ in architecture but are trained on nearly identical datasets, they are likely to produce divergent predictions when applied to scenarios that are poorly represented in the training data. This disagreement can serve as an indicator that the model's predictions in those regions are unreliable.

To quantify the uncertainty of the neural networks, we proceed as follows. We compute the evolution with each of the ten models independently with a $4^{\mathrm{th}}$ order Runge-Kutta algorithm. This gives us 10 different distribution functions for each time step. We take the mean of this distribution as the best estimate and quantify its uncertainty through a Jackknife estimate of the variance 
\begin{equation}
    \delta f(t_n) = \sqrt{\frac{M-1}{M} \sum_m \left( f_{(m)}(t_n) - \langle f(t_n) \rangle \right)^2}~,
    \label{eq:JK}
\end{equation}
where $f_{(m)}$ is the average when removing the $m$-th distribution function.

\begin{figure}
    \centering
    \includegraphics[width=0.95\linewidth]{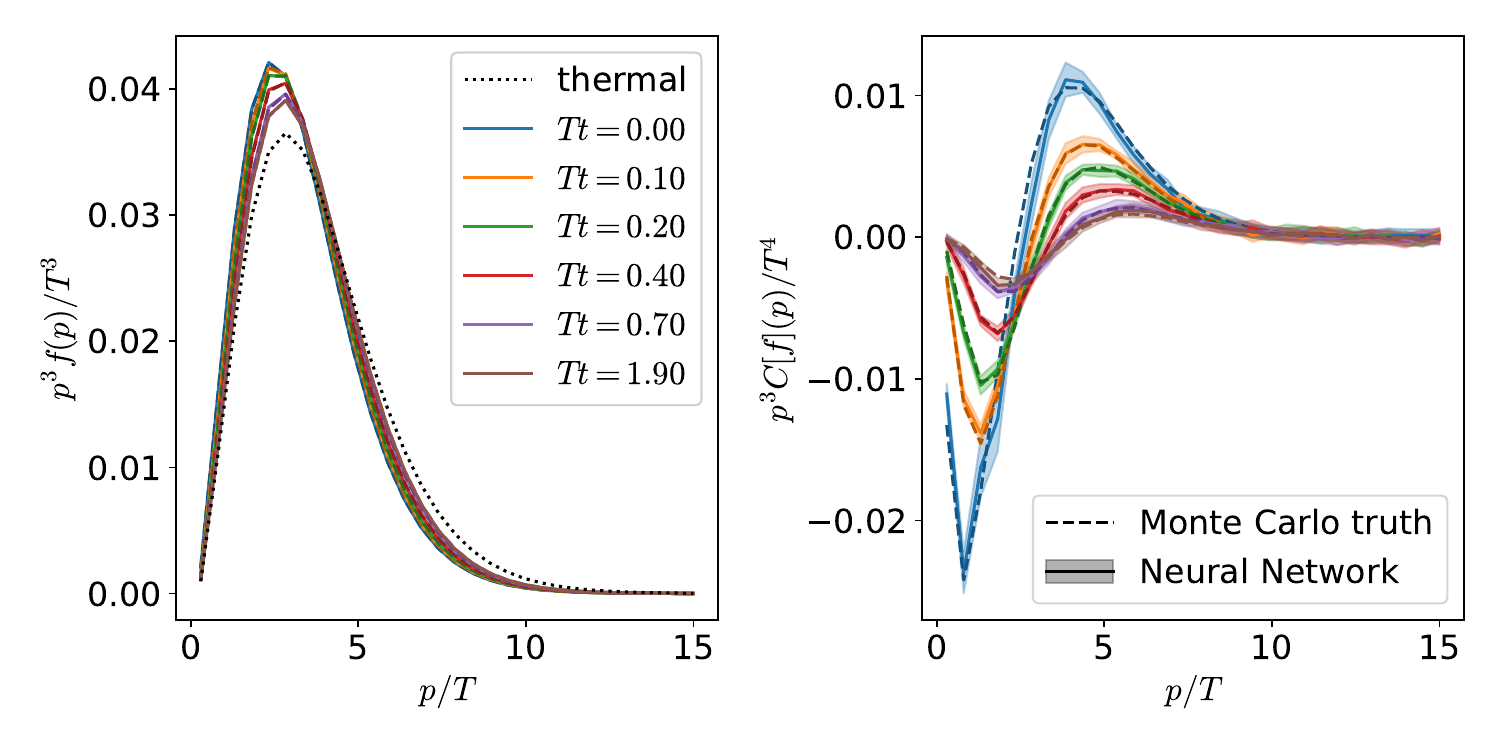}
    \includegraphics[width=0.95\linewidth]{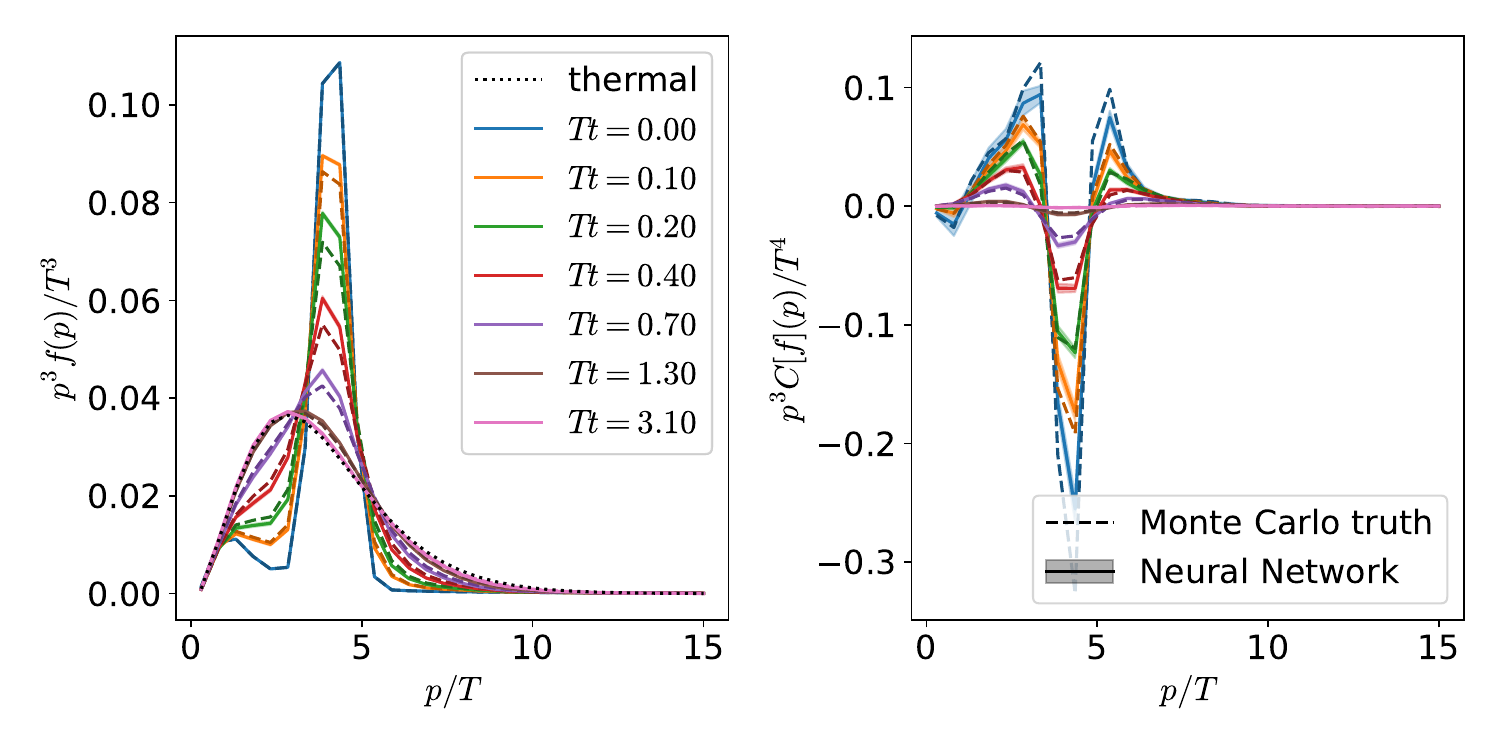}
    \includegraphics[width=0.95\linewidth]{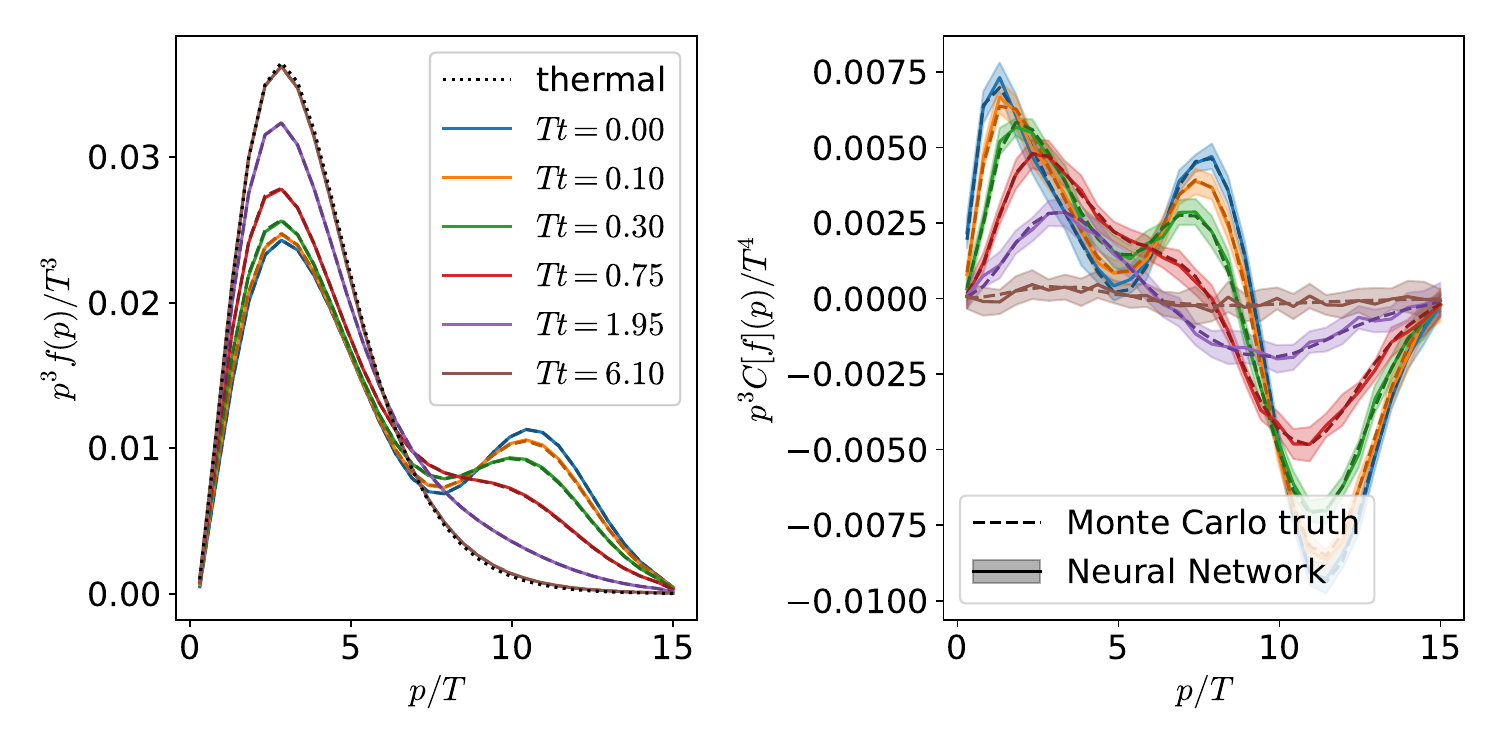}
    \caption{Three different evolutions. Left plots show the distribution functions at different times and right plots show their corresponding collision kernel with the same color. Dashed lines are the results from the Monte Carlo simulation, while the solid lines with error bars are the evolution obtained from the neural networks.
    The functional form of the initial distribution in the top two panels have not been used in the training of the network.
    }
    \label{fig:distributions_1d}
\end{figure}

\subsection{Results}

\begin{figure*}
    \centering
    \includegraphics[width=0.8\linewidth]{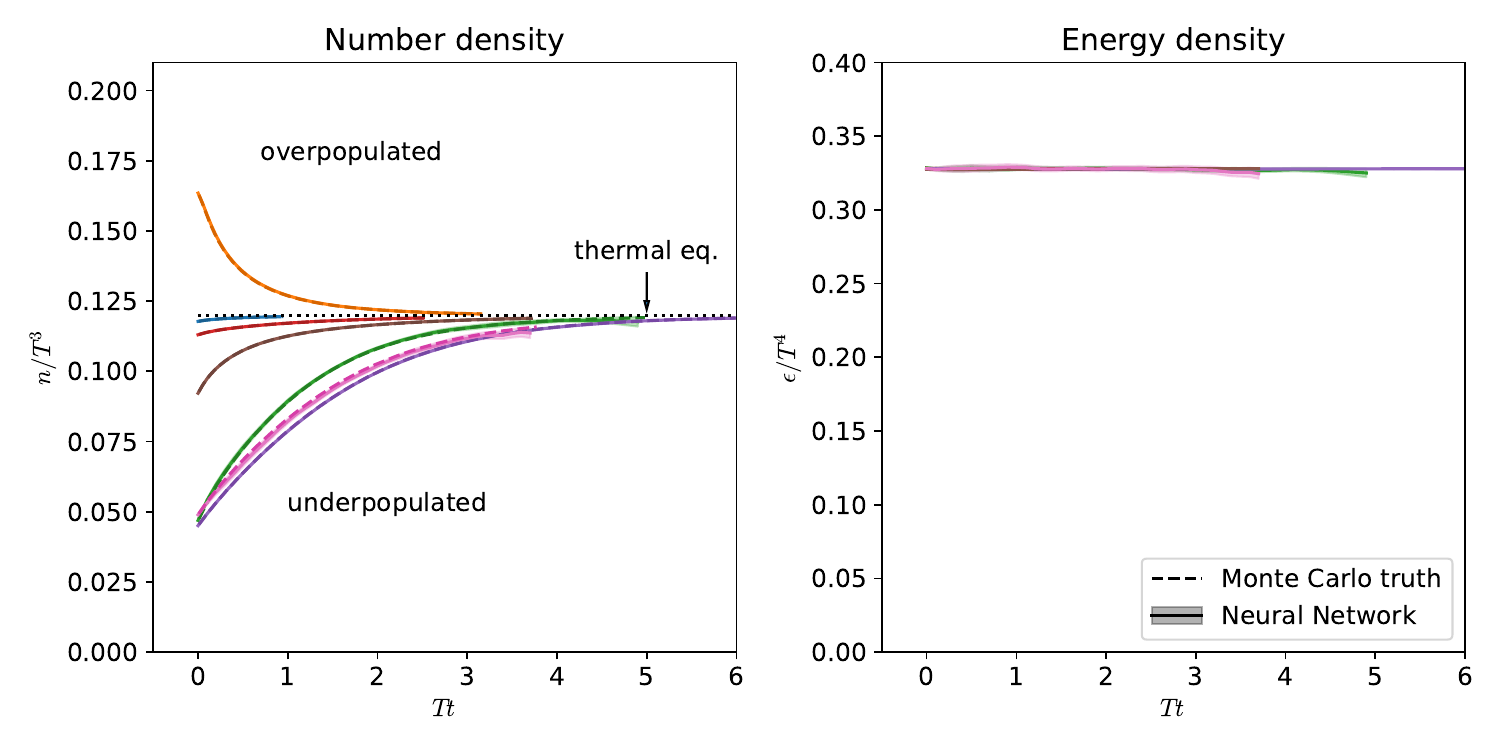}
    \caption{Number density as a function of time. Solid lines correspond to the neural network prediction, meanwhile, colored dashed lines are the EKT prediction at the same time in the evolution. Colors correspond to the same of the ones plotted in Figure \ref{fig:distributions_1d}.}
    \label{fig:number_density_1d}
\end{figure*}

The training of the 20 neural networks, 10 for each collision kernel, which we performed for this case involved around 200000 samples, which translates to a volume of data of $\sim100~\mathrm{MB}$.\footnote{We provide the data in Ref.~\cite{zenododataset}, which includes the raw data and preprocessing steps, effectively amounting to more than the $100~\mathrm{MB}$ of data.} To evaluate the performance of the described approach, we compare the evolution of several initial conditions with the results obtained within the Monte Carlo approach. In Fig.~\ref{fig:distributions_1d} we present the distribution functions and their corresponding collision kernels at different times for three different evolutions. The first two are not contained in the training data, while the third one has been extracted from the training dataset. We observe excellent agreement between the two approaches. All the Monte Carlo results lie within the error bands obtained from the neural network evolution and Eq.~\eqref{eq:JK}.

In Figure~\ref{fig:number_density_1d}, we show the number and energy density evolution for several different initial conditions over time. Again, we compare with the results of the Monte Carlo simulation, and find excellent agreement. Regarding the evolution, the number density approaches the expected thermal value, and the energy density is nicely conserved throughout the whole thermalization process. When approaching equilibrium, the error bars in the energy density increase in some of the studied cases, which signals that the network is being pushed to distributions not well represented in the training dataset. This indicates that our trained network has difficulties remaining at thermal equilibrium, which will be even more pronounced for anisotropic distributions, which we discuss in Section \ref{sec:anisotropic-distributions}.

\section{Anisotropic distributions \label{sec:anisotropic-distributions}}

We will now generalize and apply our model to anisotropic distribution functions $f(\vb p)$.

\subsection{Generation of training data, choice of initial conditions}

As in the isotropic scenario, we need an large data set to train the neural network. Now, the phase space is significantly larger due to the higher dimensionality. This needs to be considered when generating the training data, without losing sight of the fact that we aim to solve a physical problem; therefore, we should restrict ourselves to training data that is physically relevant.

Similarly as in the isotropic scenario, we generate first a set of time evolutions. These are initialized with a
generalized Kurkela-Zhu-like distribution~\cite{Kurkela:2015qoa}, where we allow for anisotropies in all three spatial directions,
\begin{equation}
    f_{\mathrm{KZ}}(p_x, p_y, p_z) = \frac{\mathcal{A}}{\lambda}\frac{1}{\tilde{p}} \exp \left[ -\frac{3}{2}\frac{\tilde p^2}{\langle p_T \rangle^2} \right]
\end{equation}
where $\tilde{p} \equiv \sqrt{(\eta p_x)^2 + (\zeta p_y)^2 + (\xi p_z)^2}$ is a squeezed momentum, $\mathcal{A}$ is a normalization constant with units of momentum and $\langle p_T \rangle \approx 1.8 Q_s$, with $Q_s$ the saturation momentum.
The data is generated such that the anisotropies $\eta, \zeta, \xi \in [0.5, 4]$ with $\xi > \zeta >\eta$ such that $P_z < P_y <P_x$, in accordance with our discussion in Section~\ref{sec:conformalsymmetry}.

In addition to the training data involving time evolutions, we add distribution functions for which we calculate the collision kernel without performing a time evolution.
This is to complement with distributions close to equilibrium to help the convergence at the thermal fixed point.
For that, we start with a thermal distribution and 
squeeze or stretch it along the radial direction controlled by additional parameter $\alpha$, $f_{\mathrm{BE}}(p) \rightarrow f_{\mathrm{BE}}(\alpha p)$, $\alpha \in [0.5, 5]$. Then, we add or subtract up to 9 three-dimensional Gaussians with random amplitudes and widths. Finally, the distribution is normalized to satisfy the condition of unit temperature $\hat{T}=1$ discussed in Section \ref{sec:conformalsymmetry}.
In summary, this set of distributions is generated by
\begin{equation}
    f(\mathbf{p}) = \frac{f_{\mathrm{BE}}(\alpha p)}{\mathcal{N}} \left[ 1 + \sum_{i=0}^{10}\!A_i \exp\! \left( \!-\!\!\!\!\sum_{j=x,y,z}\frac{(p^j - Q_i^j)^2}{2\sigma_i^j} \right) \right]
\end{equation}
with amplitudes $A_i \in[-1,1]$. The $Q_j$ are the average momentum for each of the cartesian directions, $Q_i \in [0,4Q_s]$, and the widths $\sigma_j \in [Q_s, 4Q_s]$. In general, this distribution is neither in the rest frame nor respects the pressure hierarchy $P_z < P_y <P_x$. Therefore, we additionally perform the transformation to the rest frame as discussed in Section \ref{sec:conformalsymmetry}.

\subsection{Training data}
The training for the anisotropic case consists of a total of 100000 data samples, which translates to $\sim 50~\mathrm{GB}$ of data to process. Approximately half of this data comes from static distributions, the other half from time evolutions. This emphasizes the importance of coarsening the grid as a means of data reduction. The training data was generated in a $64\times63\times63$ fine grid, while the coarse grid used for training the neural network has dimensions $32\times25\times25$, where the numbers denote the grid points in $p$, $\cos\theta$, and $\phi$, respectively. Thus, the coarsening reduces the size of the training data by a full order of magnitude.

\subsection{Neural network and hyperparameter optimization}

The hyperparameter optimization is performed similarly to the isotropic case. For the anisotropic case, we explore 500 different architectures for each collision kernel. We allow the networks to have from 1 to 3 inner layers and 64, 128, 192, 256, 386, or 512 nodes per layer. In this case, we have a $32\times25\times25$ grid; the first and last layers have a huge number of connections. To reduce the number of hyperparameters, we allow the training algorithm to perform some pruning\footnote{The pruning technique consists of setting to zero some parameters that were close to that value during the training of the neural network. With this, it is possible to reduce the size of the model we use for training, as well as its efficiency, without any important impact on the accuracy of its predictions. In our case, we combine it with the training in such a way that, after $N$ epochs, we prune a certain number of the parameters of the network and then we continue the training.} in these two layers. The frequency with which the pruning is performed and its sparsity are also treated as hyperparameters.

\subsection{Results }

\begin{figure*}
    \centering
    \includegraphics[height=0.3\linewidth]{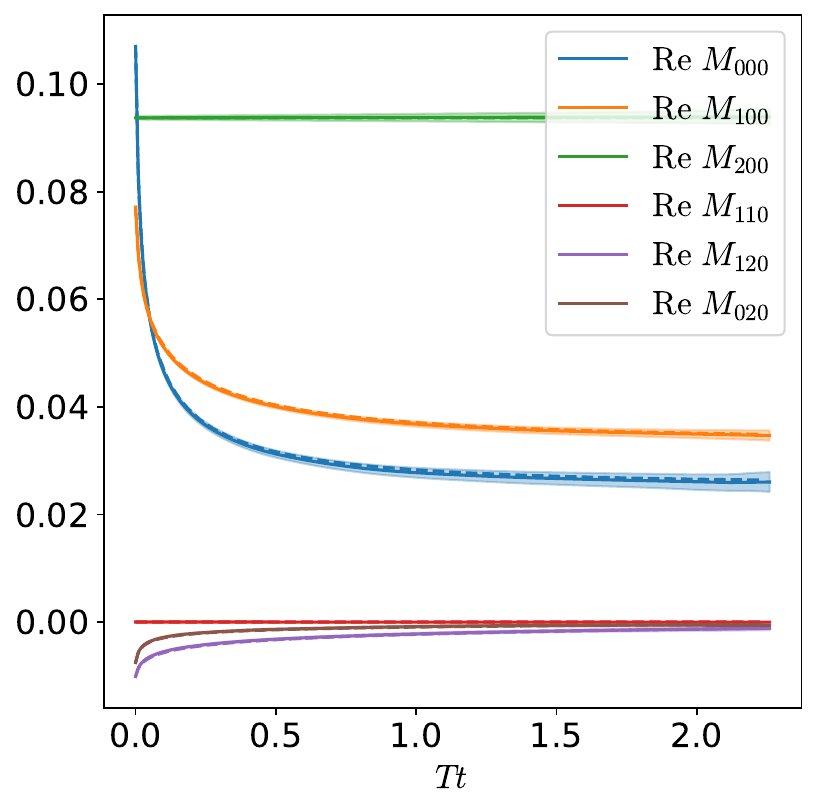}
    \includegraphics[height=0.3\linewidth]{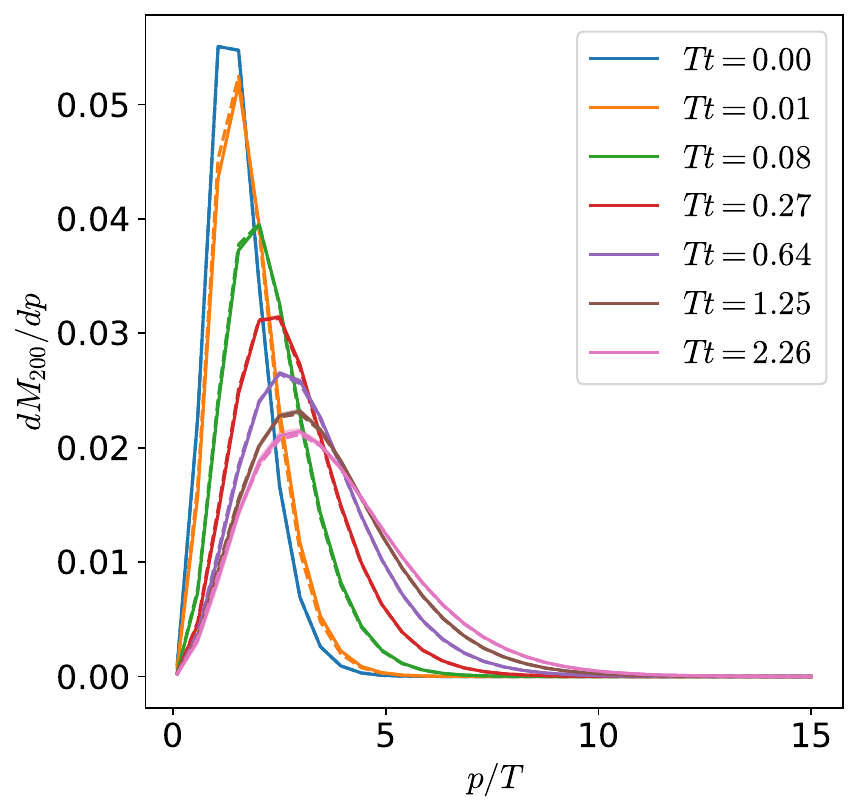}
    \includegraphics[height=0.3\linewidth]{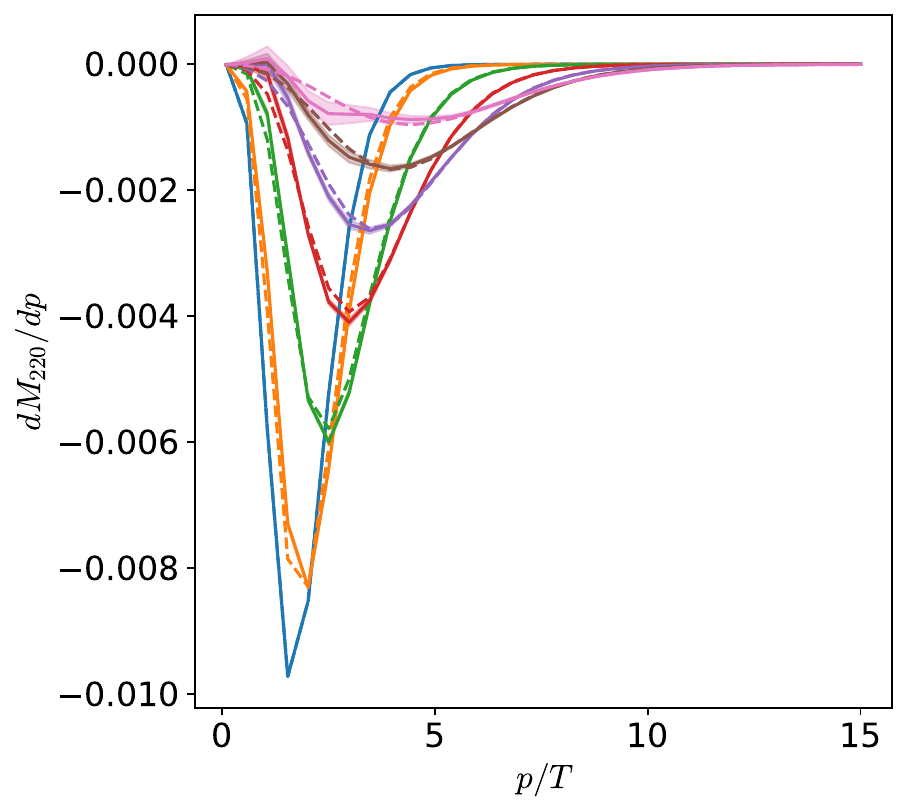}
    \includegraphics[height=0.3\linewidth]{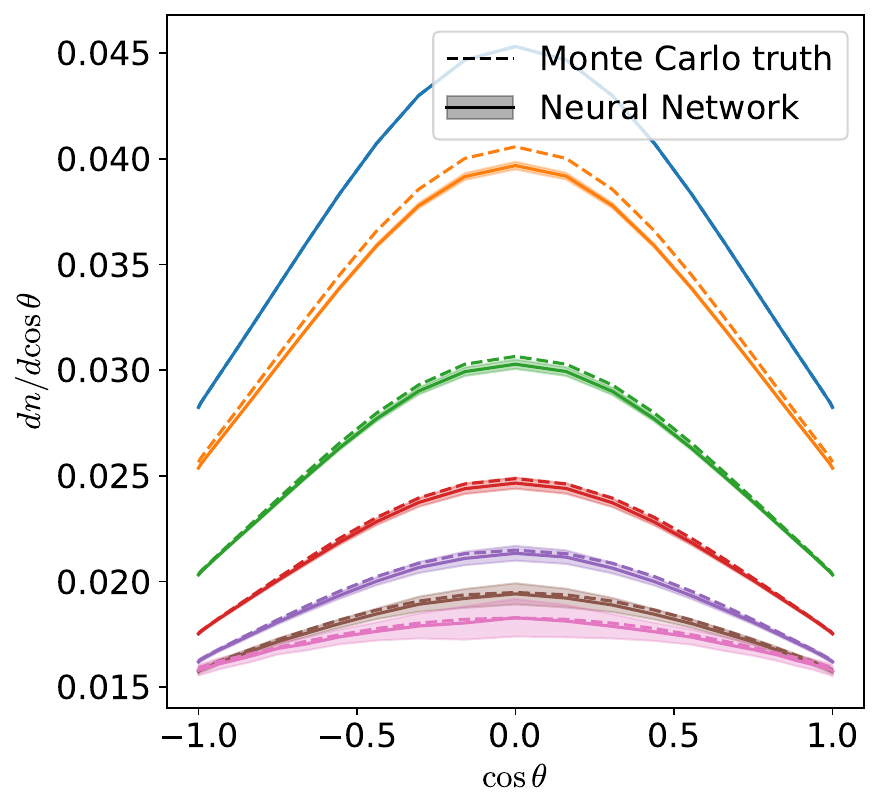}
    \includegraphics[height=0.3\linewidth]{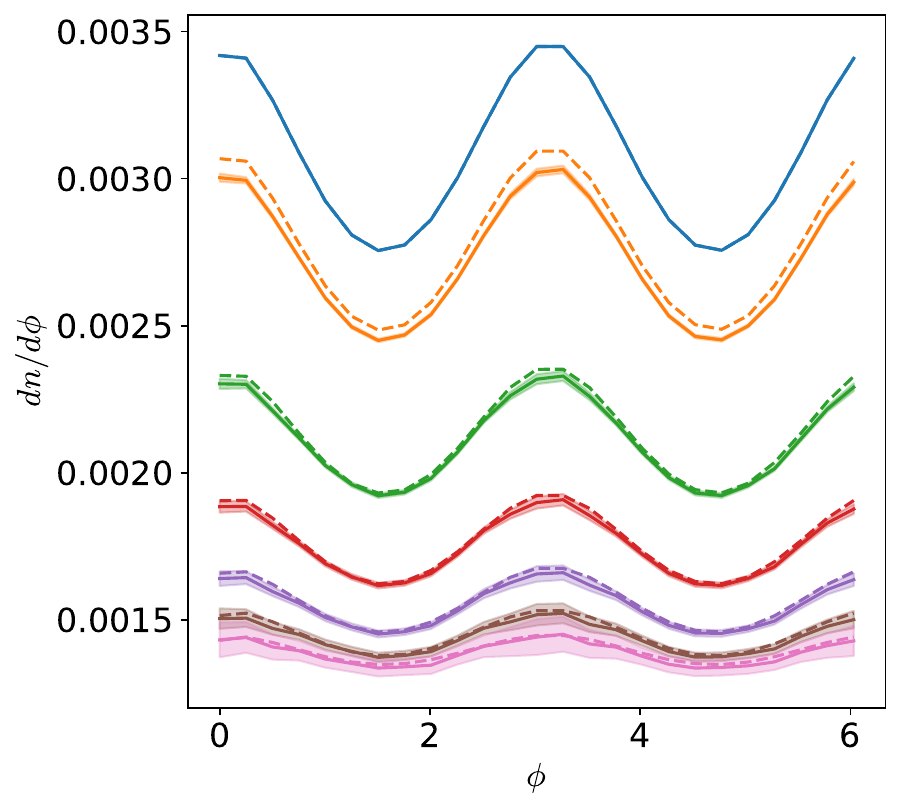}
    \caption{Time evolution for a 3D distribution. The top-left panel shows the evolution of several integral moments of the distribution as defined in Eq.~\eqref{eq:momentsadim}. The top center and top right present the $M_{200}$ and $M_{220}$ moments, respectively, differential in $p$ at different times of the evolution. Bottom panels show the moments differential in the two angular coordinates.
    }
    \label{fig:3D-differential}
\end{figure*}

\begin{figure*}
    \includegraphics[height=0.3\linewidth]{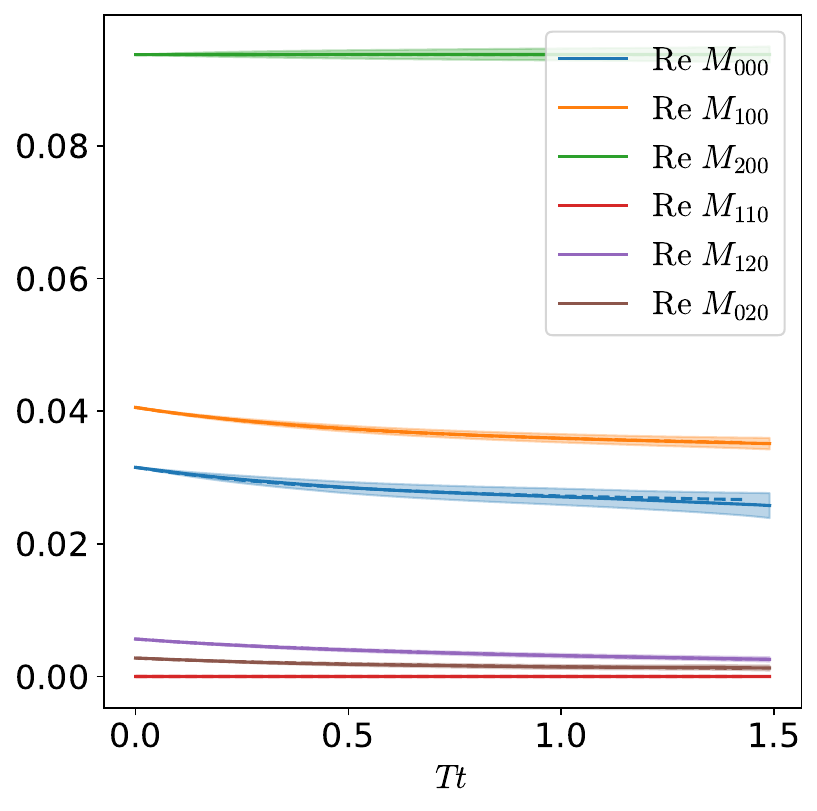}
    \includegraphics[height=0.3\linewidth]{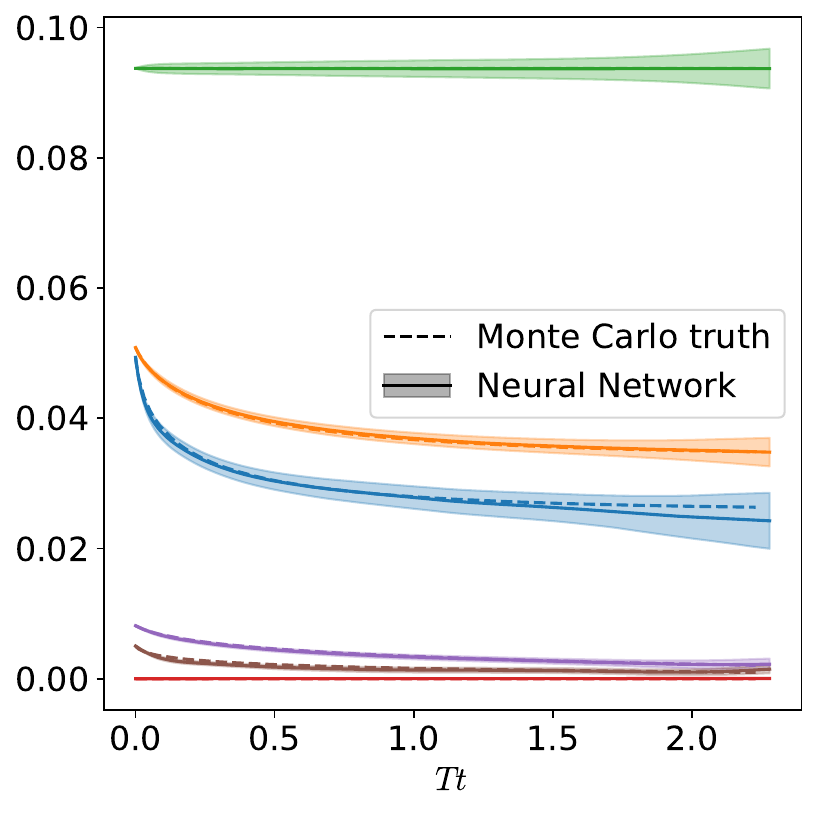}
    \includegraphics[height=0.3\linewidth]{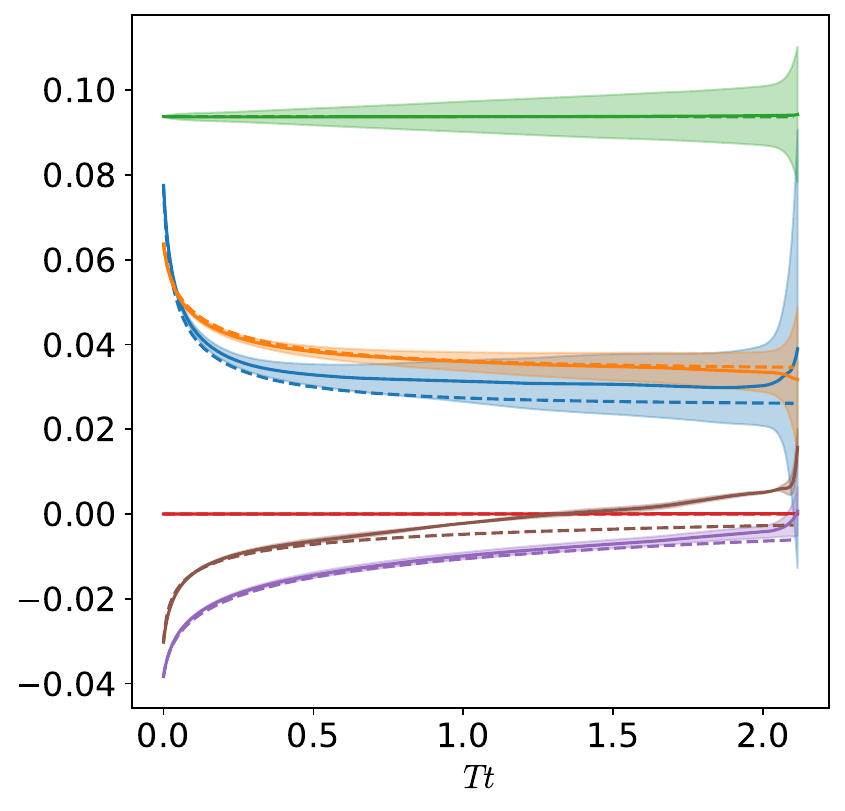}
    \caption{Time evolution of the moments defined in Eq.~\eqref{eq:momentsadim} for three different initial conditions.}
    \label{fig:Several_moments}
\end{figure*}
We use the same algorithm to compute the time evolution as that described for the isotropic case: evolve the best-performing 10 neural networks obtained in the hyperparameter tuning and use them to estimate the error of the network. We present a detailed comparison with the simulation coming from a Monte Carlo in Fig.~\ref{fig:3D-differential}. Here, we follow the time evolution of the dimensionless integral moments defined as
\begin{align}
    M_{nlm}=\frac{1}{T^{n+2}}\int\frac{\dd[3]{\vb p}}{(2\pi)^3}p^{n-1} Y^m_l{}^\ast(\theta,\phi)f(\vb p).
    \label{eq:momentsadim}
\end{align}
The first index $n\geq 0$ labels the power of $p$ in the integrand. The indices $l$ and $m$ are the indices belonging to the spherical harmonics $Y_l^m(\theta,\phi)$, and obey the condition $l \geq 0$ and $-l\leq m \leq l$. Note that the Debye mass is proportional to the moment $M_{000}$, the number density to $M_{100}$, the energy density to $M_{200}$. Other familiar quantities can also be obtained from these moments. For instance, the pressure $P_z$ is given by $P_z/T^4=\frac{1}{3}M_{200}+\frac{2}{3}M_{220}$. 

The upper left panel of Fig.~\ref{fig:3D-differential} shows the time evolution of six different moments, showing an excellent agreement between the Monte Carlo results and the predictions of the neural network. The other panels of Fig.~\ref{fig:3D-differential} show moments differential in  (or not integrated over) $p$ (upper panels) and $\cos\theta$ and $\phi$ (lower panels).
There, we observe the very good agreement of the distribution function throughout the time evolution and for different regions in phase space.
We can explicitly observe the isotropization in both angles $\theta$ and $\phi$.
It can be seen in the bottom panels that at $Tt=0.01$ the neural network slightly underpredicts the actual evolution, which is slightly above the error band.
Additionally, the growing error bars at late times indicate the cumulative error of many time steps.
Nevertheless, the neural network manages to reproduce the overall evolution very well both qualitatively and quantitatively.

We provide additional comparisons for simulations with different initial conditions in Fig.~\ref{fig:Several_moments}. There, we again first note the good overall agreement between the neural network and the Monte Carlo evolution.
However, especially in the right panel, the error bars significantly grow during the evolution. Especially close to equilibrium, the network seems to struggle to remain and approach the thermal fixed point. Thus, despite enlarging the training data with a large amount of static distributions close to equilibrium, the neural network fails to accurately learn the approach to thermal equilibrium. This indicates that even more training data might be needed, or a different network architecture that by construction conserves the thermal fixed point. A promising possibility for enlarging the training dataset would be taking the predictions of the neural network close to equilibrium, and evolve them with the Monte Carlo method. This would generate more training data which complements the first dataset close to the thermal fixed point.

Let us now provide an estimate of the performance and execution time between the Monte Carlo method and using the neural network. This is complicated by the fact that we are not using the same grid and time step. Additionally, while the Monte Carlo method reliably approaches equilibrium and is stable around it, the neural network seems to have problems staying at equilibrium, which further complicates a straightforward comparison.
Nevertheless, we can compare the time needed to perform one of the full 3D simulations with both algorithms. In the Monte Carlo approach, the time needed for the results displayed here is, on average, one day of calculations on $\mathcal O(30)$ computing cores. On the other hand, the novel neural network approach achieves the latest times of the evolution in a few minutes on a single core. That is, it exhibits a speed-up of three orders of magnitude with respect to the traditional Monte Carlo simulations.

\section{Conclusions and outlook\label{sec:conclusions}}
The deliverable of our work here is the neural network model of the collision kernel, which is publicly available \cite{zenododataset}. While we demonstrate its speed and accuracy in evolving across a variety of conditions, we note that there is room for improvement. In particular, our exploratory study is restricted to gluodynamics -- the inclusion of quarks would be straightforward \cite{Kurkela:2018wud}. Similarly simple would be the inclusion of electromagnetic probes \cite{Garcia-Montero:2023lrd}. We have not attempted to condition the network with jet-like distributions which contain sharp features \cite{Schlichting:2019abc}. It is possible that more sophisticated network architectures may be needed to capture the multi-scale physics present in such systems. 

Importantly, the major drawback of the current network is the failure to remain at the thermal fixed point. While we subtract the equilibrium distribution to have a simple zero-to-zero map, and complement the training data with distributions close to equilibrium, this does not seem to be enough to provide a stable thermal fixed point. Nevertheless, our model can be used in phenomenological applications provided that an additional stability condition is applied which guarantees the convergence towards equilibrium.

In the future, diffusion models might be useful for the generation of the collision kernel, or for generating training data to further improve the neural network. These models are very prominently used in image generation, and now also venture into the domain of heavy-ion physics \cite{Sun:2024lgo}.

With the neural network in hand, the obvious next step is to implement the EKT collision kernel into an existing kinetic-theory framework capable of handling 3+1D advection \cite{Kurkela:2020wwb,Taghavi:2025xhl,Ambrus:2024hks}. It remains to be seen whether the training dataset used here is sufficiently generic to accurately support the time evolution in phenomenologically relevant systems. Regardless of the outcome, we have demonstrated the usefulness of the neural network approach to EKT. If fully 3+1D simulations require new training data, the simulations themselves will provide the necessary distribution functions that we can use to construct further training data using the methods developed here. 

\begin{acknowledgments}
We would like to thank Andreas Ipp, David I.~Müller, and Kayran Schmidt for insightful discussions.

FL is supported by the Austrian Science Fund (FWF) under Grant DOI 10.55776/P34455. FL is a recipient of a DOC Fellowship of the Austrian Academy of Sciences at TU Wien (project 27203). The results in this paper have been achieved in part using the Austrian Scientific Computing (ASC) infrastructure, project 71444. SBC is supported by the European Research Council project ERC-2018-ADG-835105 YoctoLHC; by Mar\'\i a de Maeztu grant CEX2023-001318-M and by project PID2023-152762NB-I00, both funded by MCIN/AEI/10.13039/-501100011033; from the Xunta de Galicia (CIGUS Network of Research Centres) and the European Union. SBC also acknowledges the support of Axudas de apoio \'a etapa predoutoral program (Ref. ED481A 2022/279).

\end{acknowledgments}

\appendix
\section{Numerical details on the Monte Carlo evaluation of the collision kernels\label{sec:collision-kernels}}
\subsection{Explicit formulas for elastic and inelastic collision kernels}
For a numerical evaluation, it is convenient to rewrite the elastic collision kernel Eq.~\eqref{eq:c22_first} in a symmetric form,
\begin{align}
\Ctwotwo&[f(\tilde{\vb{p}})] = \frac{\left( 2 \pi \right)^3}{4 \pi \tilde{p}^2} \dfrac{1}{8 \nu} \int \dd{\Gamma_{\mathrm{PS}}} \left|\mathcal M(\vb p,\vb k;\vb p', \vb k') \right|^2  \nonumber\\ \label{eq:C22}
& \times  \Big( f(\vb{p}) f(\vb{k}) (1 + f(\vb{p^\prime})) ( 1 + f(\vb{k^\prime}) \\
&\qquad - f(\vb{p^\prime}) f(\vb{k^\prime}) (1 + f(\vb{p}) (1 + f(\vb{k}) \Big)  \nonumber\\
& \times \left( \delta^3(\vb{\tilde{p}} - \vb{p})  + \delta^3(\vb{\tilde{p}} - \vb{k}) - \delta^3 (\vb{\tilde{p}} - \vb{p^\prime}) - \delta^3(\vb{\tilde{p}} - \vb{k^\prime})\right), \nonumber
\end{align}
with the integration measure
\begin{align}\label{eq:integration-measure}
 \int \dd{\Gamma_{\mathrm{PS}}}  &=  \int_{\vb{p k p' k'}}
 \left(2 \pi \right)^4 \delta^4\left( P + K - P^\prime - K^\prime \right) \\
 & = \frac{1}{2^{12} \pi^8} \int_0^\infty \dd q \int_{-q}^q \dd \omega \int_{\frac{q-\omega}{2}}^\infty \dd p  \int_{\frac{q+\omega}{2}}^\infty \dd k \nonumber \\ 
 & \times \int_{-1}^1 \dd \cos\theta_q\int_0^{2\pi}\dd{\phi_q} \int_0^{2 \pi} \dd \phi_{pq} \int_0^{2 \pi} \dd\phi_{kq}, \nonumber
 \end{align}
The integration variables $p=|\vb p|$ and $k=|\vb k|$ denote the length of the corresponding three-vector. The vector $\vb q=\vb p' - \vb p$ is the exchanged momentum, $\omega=p'-p$ denotes the exchanged energy. We first perform the $\vb q$ integral (parametrized by its spherical coordinates $q$, $\theta_q$, and $\phi_q$) and then perform the $\vb p$ and $\vb k$ integral in a frame, in which $\vb q$ points in the $z$ direction. In these frames, the polar angles $\theta_{pq}$ and $\theta_{kq}$ are kinematically fixed (see, e.g., \cite{Arnold:2003zc}), and only the azimuthal angles $\phi_{pq}$ and $\phi_{kq}$ remain.

The integrand is symmetric under the exchange of the outgoing particles with momenta $k'\leftrightarrow p'$. This exchanges the $u-$ and $t-$channel, and shifts the dominating integration region from small $|u|$ and small $|t|$ to the small $|t|$ region only, greatly simplifying the screening prescription and importance sampling. 

The matrix element for gluon-scattering is given by
\begin{align}
\label{eq:ggMatrixElem}
\frac{\left|\mathcal M\right|^2}{4 \lambda^2 d_A} = 9 + \frac{(s-t)^2}{\underline{u^2}} + \frac{(u-s)^2}{\underline{t^2}} + \frac{(t-u)^2}{s^2},
\end{align}
where the gauge coupling and number of colors conveniently combine to the 't Hooft coupling $\lambda=g^2\Nc$. The underlined terms need to be modified to include self-energy corrections \cite{Arnold:2002zm}
\begin{align}
	{\frac{(s-u)^2}{\underline{t^2}}}&\to \left|G^{\mathrm{ret}}_{\mu\nu}(P-P')\;(P+P')^\mu(K+K')^\nu\right|^2\label{eq:amy_replacement},
\end{align}
where $G_{\mu\nu}^{\mathrm{ret}}$ is the retarded gluon propagator, which we use in an isotropic HTL (isoHTL) approximation. More details on the screening can be found in Ref.~\cite{Boguslavski:2024kbd}.

The inelastic collision term accounts for collinear splitting and merging. In its symmetrized form, it is given by
\begin{align}
    \begin{split}\label{eq:c12}
        \Conetwo[f(\vb {\tilde p})]&=
        \frac{(2\pi)^3}{4\pi\tilde p^2}\frac{1}{\nu}\int_0^\infty \dd{p}\int_0^{p/2}\dd{k'}4\pi\gamma^{p}_{p',k'}\\
        &\quad\times\Big\{f(\vb p)(1+f(p'\vbphat))(1+f(k'\vbphat))\\
        &\qquad\quad-f(p'\vbphat)f(k'\vbphat)(1+f(\vb p))\Big\}\\
        &\times\left[\delta(\tilde p-p)-\delta(\tilde p-p')-\delta(\tilde p-k')\right]
    \end{split}
\end{align}
with $p'=p-k'$ and $\vbphat = \vb p / p$ being a unit vector, which enforces that all particles are collinear in the direction of $\vb p$.
The effective splitting/joining rates $\gamma^p_{p'k}$  interpolate between the Bethe-Heitler and LPM regime and are obtained via \cite{Arnold:2002zm}
\begin{subequations}
\begin{align}
    \gamma^{p}_{p',k}=\frac{p'^4+p^4+k^4}{p'^3p^3k^3}\mathcal F^{\vbphat}(p,p',k),
\end{align}
with $\mathcal F$ given by
\begin{align}
    \mathcal F^{\vbphat}(p',p,k)=\frac{\lambda\nu}{8(2\pi)^3}\int\frac{\dd[2]{\vb h}}{(2\pi)^2}2\vb h\cdot \mathrm{Re}\, \vb F(\vb h).
\end{align}
Here, $\mathrm{Re}\,\vb F(\vb h)$ is the real part of the solution to the integral equation
\begin{align}
\begin{split}
    2\vb h &= i\delta E(\vb h; p',p,k)\vb F(\vb h)\\
    &+\frac{\lambda T_\ast}{2}\int\frac{\dd[2]{\vb \qperp}}{(2\pi)^2}\left(\frac{1}{\qperp^2}-\frac{1}{\qperp^2+m_D^2}\right)\\
    &\times\Big\{3 \vb F(\vb h)-\vb F(\vb h-k\vb \qperp)-F(\vb h-p'\vb \qperp)-F(\vb h-p\vb \qperp)\Big\}\label{eq:integral_equation}
    \end{split}
\end{align}
and 
\begin{align}
    \delta E(\vb h; p',p,k)=\frac{m_D^2}{4}\left(\frac{1}{k}+\frac{1}{p}+\frac{1}{p'}\right)+ \frac{\vb h^2}{2pkp'}.
\end{align}
\end{subequations}
In writing Eq.~\eqref{eq:integral_equation}, we have implicitly taken an isotropic screening approximation, where the Wightman correlator of the gluon field generated by the other particles in the plasma can be simplified using a sum rule \cite{Aurenche:2002pd, Arnold:2002zm}.
The effective infrared temperature $T_\ast$ is given by
\begin{align}
    T_\ast = \frac{2\lambda}{m_D^2}\int\frac{\dd[3]{\vb p}}{(2\pi)^3}f(\vb p)(1+f(\vb p)),\label{eq:Tstar}
\end{align}
and the Debye mass is defined as
\begin{align}
	m_D^2 = 8\lambda\int\frac{\dd[3]{\vb p}}{(2\pi)^32|\vb p|}f(\vb p).\label{eq:debye_mass_general}
\end{align}

\subsection{Monte Carlo evaluation of the collision terms}
Rewriting the Boltzmann equation \eqref{eq:boltzmann_equation} in terms of the number moments $n_{ijk}$,
\begin{align}
    v^\mu\partial_\mu n_{ijk}&=\int\frac{\dd[3]{\vb {\tilde p}}}{(2\pi)^3}w_{ijk}(\vb {\tilde p})v^\mu\partial_\mu f(\vb{\tilde p})\\
    &=-\int\frac{\dd[3]{\vb {\tilde p}}}{(2\pi)^3}w_{ijk}(\tilde{\vb p})\mathcal C[f(\vb{\tilde p})]
\end{align}
replaces the delta functions in the collision terms \eqref{eq:C22} and \eqref{eq:c12} by the wedge functions $w_{ijk}$.
Here, we have used the abbreviation $w_{ijk}(\vb p) = w_i(p)  \tilde w_j(\cos\theta)\hat w_k(\phi)$.
The collision terms are then computed using Monte Carlo importance sampling.

For the elastic collision term $\Ctwotwo$, points from the integration measure \eqref{eq:integration-measure} are sampled using Monte Carlo importance sampling. Having finite grid boundaries, we need to require that all momenta sampled lie within the grid.
For the $q$, $\omega$, $p$, and $k$ integration, the grid boundaries $\pmin$ and $\pmax$ have to be considered, and
these integrals become
\begin{align}
    &\int_0^{\pmax}\dd{q}\int_{\max(-q,\pmin-\pmax)}^{\min(q,\pmax-\pmin)}\dd{\omega}\\
    &\int_{\max\left(\frac{q-\omega}{2},\pmin,\pmin-\omega\right)}^{\min(\pmax,\pmax-\omega)}\dd{p}\int_{\max\left(\frac{q+\omega}{2},\pmin,\pmin+\omega\right)}^{\min(\pmax,\pmax+\omega)}\dd{k}\nonumber,
\end{align}
which ensures that all momenta $k$, $k'=k-\omega$, $p$, $p'=p+\omega$ lie within the grid.
The second boundaries for $\omega$ come from the requirement that $\pmax>\pmin+\omega$ and $\pmax+\omega > \pmin$ in the $k$ integral.

For the $q$-integral, we sample from a $\dd{q}/(q+\xi m_D)^4$ distribution, with $\xi=e^{5/6}/\sqrt{8}$ chosen such that this approximates well the full hard thermal loop result, although its precise value is not important for the sampling process. While $\omega$ is sampled uniformly, both $k$ and $p$ are sampled from a $\dd{k}/k$ distribution.

For the inelastic collision term, $p$ is sampled from a $\dd{p}/(p+\xi m_D)$ distribution, and $k'$ is sampled from a $\dd{k'}/{k'^2}$ distribution.

\section{Restframe\label{app:restframe}}
The rest frame is defined as the frame in which the energy-momentum tensor given by Eq.~\eqref{eq:tmunu},
\begin{align}
    T^{\mu\nu}=\nu\int\frac{\dd[3]{\vec p}}{(2\pi)^3}\frac{p^\mu p^\nu}{p^0} f(\vec p)=\varepsilon u^\mu_0 u^\nu_0 
    + \sum_i P_i u^\mu_i u^\nu_i,
\end{align}
is diagonal. It can be decomposed in its eigenvectors
$\{u_\nu^\mu \}$ (lower index labeling the eigenvectors), where its diagonal components give the energy density $\varepsilon$ and the pressures $P_i$ in different directions. Importantly, $u_0$ is the only timeline eigenvector, which is normalized to $u_0^2=-1$, while the others are spacelike $u_i^2=1$. Note that we use the mostly-plus metric convention here, $\eta_{\mu\nu}=\mathrm{diag}(-1,1,1,1)$.
Diagonalizing a matrix is a well-known problem in linear algebra.
In our case, the eigenvector equation is given by
\begin{align}
    T^\mu{}_\nu u^\nu_i = \lambda_i u^\mu_i.
\end{align}
It is a basic fact in linear algebra that $T^{\mu}{}_{\nu}$ is diagonalized by the matrix consisting of the eigenvectors, i.e.,
\begin{align}\label{eq:Lorentzfrafodiag}
    P^{-1}{}^\mu{}_\alpha T^\alpha{}_\beta P^\beta{}_\nu = \begin{pmatrix}
        \lambda_1 & 0 & 0 & 0\\
        0 & \lambda_2 & 0 & 0 \\
        0 & 0 &\lambda_3 & 0 \\
        0 & 0 & 0 & \lambda_4
    \end{pmatrix}
\end{align}
and $P$ is given by the eigenvectors inserted in every column as a column vector,
\begin{align}
    P^\mu{}_\nu = \begin{pmatrix}
        u_0^\mu & u_1^\mu & u_2^\mu & u_3^\mu
    \end{pmatrix} = u^\mu_i.
\end{align}
It can be easily shown that this basis change $P$ is the Lorentz transformation that transforms to the eigensystem, or to the rest frame.
The defining property of a Lorentz transformation is the invariance of the metric,
\begin{align}
    P^\mu{}_\alpha\eta_{\mu\nu} P^\nu{}_\beta = \eta_{\alpha\beta}.
\end{align}
Inserting for $P^\mu{}_\nu = u^\mu_\nu$ (lower index labeling the eigenvectors), we obtain
\begin{align}
    P^\mu{}_\alpha\eta_{\mu\nu} P^\nu{}_\beta = (u_\mu\cdot u_\nu) = \eta_{\mu\nu},
\end{align}
which shows that $P$ is a Lorentz transformation that takes us from a general frame to the rest frame.

We order the space-like eigenvectors such that $P_z < P_y < P_x$, in accordance with the discussion in Section~\ref{sec:conformalsymmetry}. This corresponds to an additional rotation of the system such that the ``most anisotropic'' direction is the $z$-direction.

However, this Lorentz transformation is not unique. This is because an eigenvector is only defined up to a multiplicative constant, and even a normalized eigenvector can still be multiplied by a factor $(-1)$. We will now discuss how to fix the transformation uniquely. First, we require that $P^\mu{}_\nu$ is an orthochronous Lorentz transformation, which means that it does not flip time (time reversal). The condition for that is that $P^0{}_0 > 0$. If $P^0{}_0 <0$,  we multiply the timelike eigenvector with a factor $(-1)$, or, equivalently, perform
\begin{align}
    P\to PT, && \mathrm{with} && T = \mathrm{diag}(-1,1,1,1).
\end{align}
After that, one might additionally perform a parity transformation to enforce positive determinant, $\det P \neq +1$. For that, one could multiply all spacelike eigenvectors with $(-1)$, or, equivalently, perform the parity transformation
\begin{align}
    P\to P\mathcal P, &&\mathrm{with} && \mathcal P = \mathrm{diag}(1,-1,-1,-1).
\end{align}
However, this would still not determine the transformation uniquely, because one may still perform rotations around any spatial axis with angle $\pi$, or, equivalently, multiply $P$ by
\begin{align}
    \mathcal R = \mathrm{diag}(1,\pm 1, \pm 1, \pm 1)\label{eq:rotation}
\end{align}
with exactly two negative signs.

Instead of the parity transformation, we make the frame choice unique by requiring that the current density
\begin{align}
    J^\mu = \int\frac{\dd[3]{\vb p}}{(2\pi)^3}\frac{p^\mu}{p^0}f(\vb p)\label{eq:current}
\end{align}
is positive for all spatial directions. This can be achieved by fixing the signs in \eqref{eq:rotation}, or, equivalently, multiplying every basis vector by $(-1)$ that corresponds to a direction of negative particle current.
Clearly, the current \eqref{eq:current} transforms Lorentz covariantly, because of the Lorentz invariant phase space measure $\dd[3]{\vb p}/p_0$ and the Lorentz vector $p^\mu$. As is evident from \eqref{eq:Lorentzfrafodiag}, for upper indices we need to use $P^{-1}$. Thus, for any Lorentz transformation $P$, we can easily obtain $J^\mu$ in the rest frame.

In practice, in our implementation, we obtain the distribution function in the rest frame 
$f_{\mathrm{LRF}}(\vb q)=\left.f\left((\Lambda^{-1})^i{}_\alpha q^\alpha\right)\right|_{q^0=|\vb q|}$ by applying the Lorentz transformation to every grid point in the rest frame and linearly interpolating on the old grid.

\bibliography{bibliography.bib}

\end{document}